\documentclass[notitlepage,aps,amssymb,showpacs,superscriptaddress,nofootinbib]{revtex4-1}
\usepackage{graphics}
\usepackage{epstopdf}
\usepackage{enumerate}
\usepackage{pstool}
\usepackage{subfigure}
\usepackage[misc,geometry]{ifsym}

\begin{document}
\title{A Study of Correlations in the Stock Market}
\author{Chandradew Sharma}
\email{csharma@goa.bits-pilani.ac.in}
\affiliation{Department of Physics, BITS Pilani K.K Birla Goa Campus, N.H. 17B Zuarinagar, Goa 403726, India.}
\author{Kinjal Banerjee}
\thanks{Corresponding Author}
\email{kinjalb@gmail.com}
\affiliation{Department of Physics, BITS Pilani K.K Birla Goa Campus, N.H. 17B Zuarinagar, Goa 403726,
India.}
\date{\today}
\begin{abstract}

We study the various sectors of the Bombay Stock Exchange(BSE) for a period of 8 years from April 2006 - March 2014. Using the data
of daily returns of a period of eight years we make a direct model free analysis of the pattern of  the sectorial indices movement 
and the correlations among them. Our analysis shows significant auto correlation among the individual sectors and also strong cross-correlation 
among sectors. We also find that auto correlations in some of the sectors persist in time. This is a very significant
result and has not been reported so far in Indian context
These findings will be very useful in model building for prediction of price movement of equities, derivatives and portfolio management. 
We show that the Random Walk Hypothesis is not applicable in modeling the Indian
market and  Mean-Variance-Skewness-Kurtosis based portfolio optimization might be required. We also find that almost all sectors are highly correlated during 
large fluctuation periods and have only moderate correlation during normal periods.    
\end{abstract}

\maketitle

%%%%%%%%%%%%%%%%%%%%%%%%%%%%%%%%%%%%%%%%%%%%%%%%%%%%%%%%%%%%%%%%%%%%%%%%%%%%%%%%%%%%%%%%%%%%%%%%%%%%%%%%%%%%%%%%%%%%%%%%%%%%%%%%%%%%%%%%%%%%%%%%%%%%
\section{Introduction}

The stock market is an extremely complex system with various interacting components \cite{Mantegna99}. 
The movement of stock prices are somewhat interdependent as well as dependent on a wide multitude of external stimuli like announcement 
of government policies, change in interest rates, changes in political scenario, announcement of quarterly results by the listed companies 
and many others. The overall result is a chaotic complex system which has so far proved very difficult to analyze and
predict. In fact it is still not completely clear, what are the generic features that will appear in any stock market
and what are the features which depend on the social, political and economic climate of the country and/or of the world.
So it is important to study each market individually so that finally we can be sure that certain behaviors or patterns
are universal. Although some amount of work has been done in understanding the stock markets in Europe \cite{lux} and the United
States\cite{USA}, the proper mathematical and statistical study of emerging markets like India are in their infancy
\cite{sitabhra1}.
 
So far, there is no exact understanding on which external stimulus has how much effect on the stock prices or even how the self interactions 
of the various stocks or the various sectors drive the market. Broadly speaking, the price movement of a particular stock can be classified as 
(i) market (common to all stocks), (ii) sector (related to a particular business sector) and 
(iii) idiosyncratic (limited to an individual stock). While it is virtually impossible to develop any theory for this idiosyncratic movement,
it is possible to analyze, study and build models for the other two types of stock movement. 
From an investors point of view, the most important reason to understand the stock market is to get the maximum 
possible return on an investment with the minimum possible risk. So a better understanding of the stock market will lead
to better theories of portfolio management.

One important step in improving our understanding of the stock market is to study how the stock price movement of one stock affects the price
of other stocks. One way to do this would be to see how one stock movement affects the others within the same sector. Another is to
study how the overall prices of the various sectors are correlated. The goal of this study is to try to determine and quantify,
from the available data, some of the possible correlations which might exist between the stock prices. This will not only enhance the
understanding of the stock market as a whole but will play a crucial role in investment decisions like portfolio management. A systematic model
independent analysis of the data that we do will also help in building more efficient and enhanced models which will give adequate weightage to
the various relations which exist between movement of stock prices across sectors in a market and may help in forecasting future trends. 
Studies of such correlations have been carried out to a limited extent in the context of New York Stock Exchange \cite{NYSE}, 
but to the best of our knowledge, no such study exists for the Bombay Stock Exchange (BSE) \cite{BSE}. 

To understand the financial market, it is very important to know the distribution of the return on a stock. 
Our data consists of the daily returns of 12 sectors of stocks of the BSE for Financial Year(FY) 2006 to FY
2013 i.e. $1990$ days from 3rd April 2006 to 31st March 2014. We will be treating each sector as one entity in the rest
of the paper. This approach is novel and has not been carried out before, at least in the context of Indian markets.
 
\begin{figure}
\centering
\subfigure[Variation of BSE index over time]{
\includegraphics[height=3in,width=3in,angle=0]{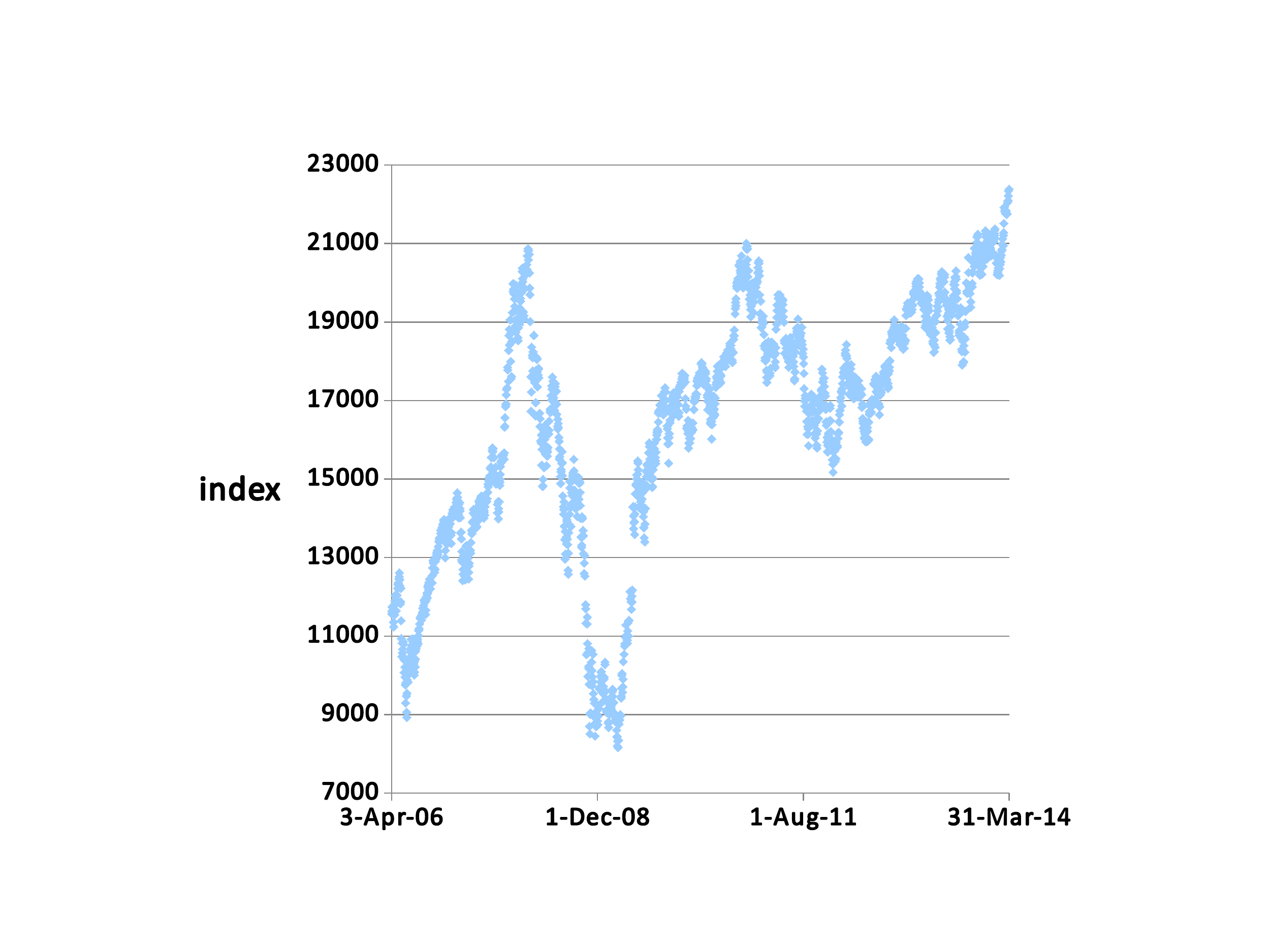}}  
\quad
\subfigure[Variation of (log) return of BSE over time]{
\includegraphics[height=3in,width=3in,angle=0]{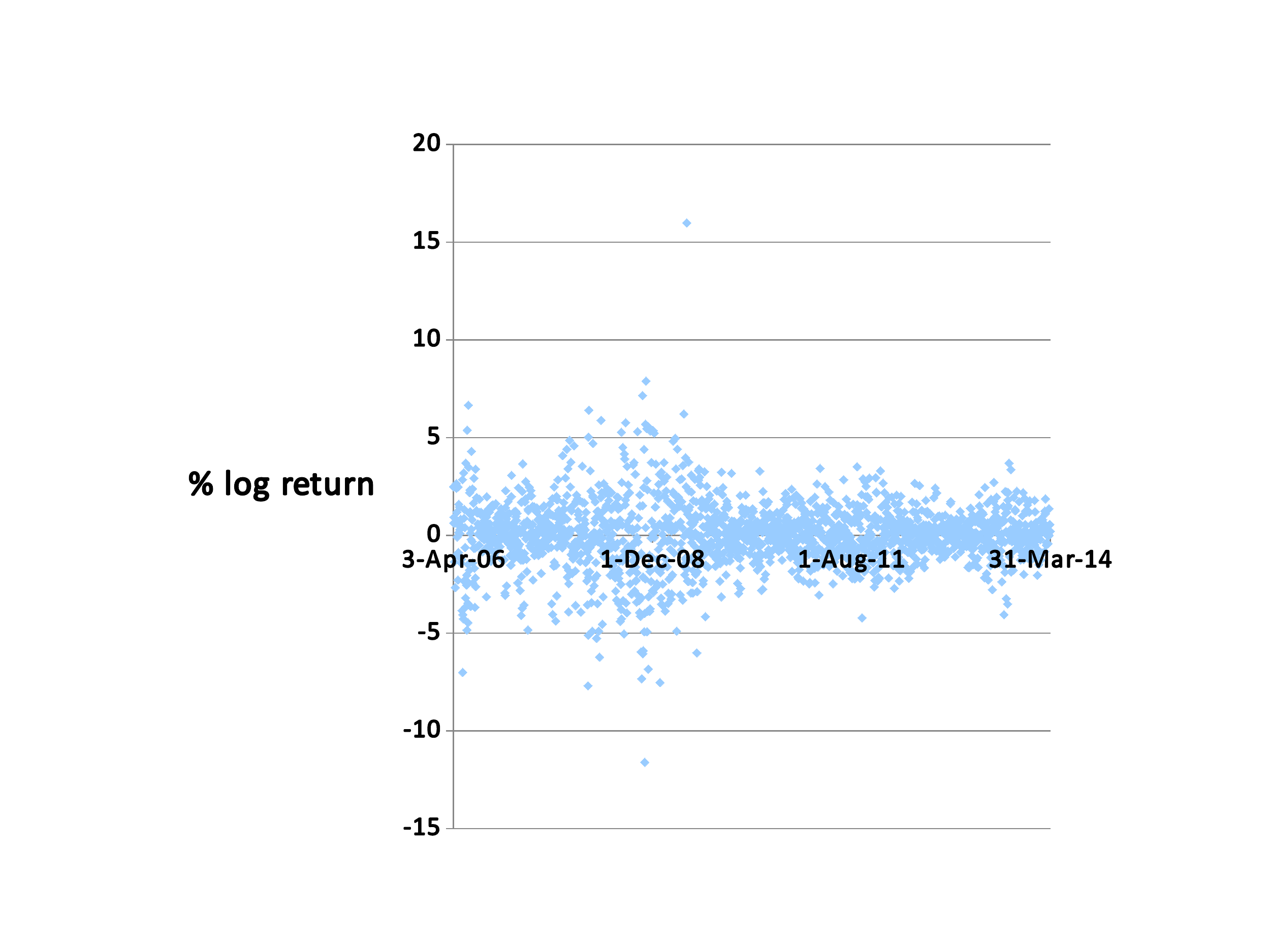}}  
\quad
\subfigure[Variation of (log) return of BSE over time]{
\includegraphics[height=3in,width=3in,angle=0]{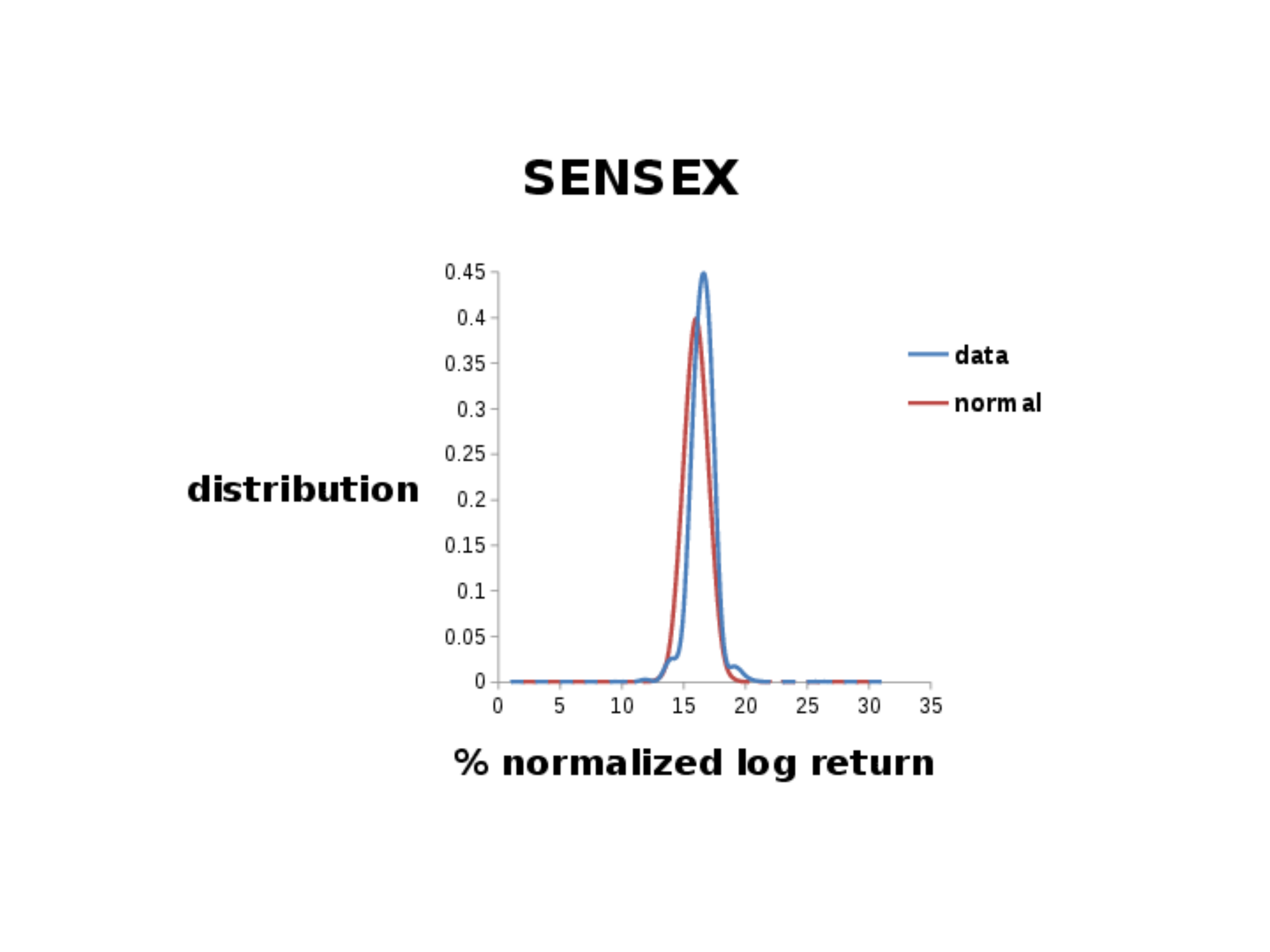}} 
\caption{Behavior of S\& P BSE SENSEX for April 2006-Mar 2014}
\label{figure1}
\end{figure}

If $P_{i}(t)$ is the index of the sector $i=1,\dots,N$ at time $t$, then the \textit{(logarithmic) return} of the $i$th sector over a
time interval $ t = 1 $ to $ t= T $ days in the interval is defined as
\begin{equation}
R_{i}(t) \equiv \ln {P_{i}(t+1)}- \ln {P_{i}(t)} \label{return}
\end{equation}
In our case $T=1900$, the number of days we have considered, and $N =13$ because we look at the following 
12 sectors S$\&$P BSE Auto (Auto), S$\&$P BSE Bankex (Bankex), S$\&$P BSE Consumer Durables (CD), S$\&$P BSE Capital
Goods (CG) , S$\&$P BSE FMCG (FMCG), S$\&$P BSE Health care (HC), S$\&$P BSE IT (IT), S$\&$P BSE Metal (Metal), 
S$\&$P BSE Oil and Gas (Oil and Gas), S$\&$P BSE Power (Power), S$\&$P BSE Realty (Realty) 
and S$\&$P BSE Teck (Teck) and the S$\&$P BSE SENSEX (Sensex) which serves as the benchmark. The plot of the Sensex index and the log return 
over the time interval under consideration is given in Figure \ref{figure1}.  From Figure \ref{figure1}, 
it is clear that we can divide the entire period in two sub interval 
(i) from FY 2006 - FY 2009 as large fluctuation period and (ii) from FY 2010 - FY 2013 for normal period. We shall discuss
how the cross correlations of the sectors are markedly different in these two periods, later in the paper. 
Obviously the mean return of the $i$th sector is given by
\begin{equation}
\bar {R_{i}}  = \frac{1}{T}\sum \limits_{t=1}^T R_{i}(t)
\end{equation}
Defining $R^{\prime}_{i}  = (R_{i}(t)-\bar {R_{i}})$, we can write the $k$th moment of the $i$th sector as
\begin{equation}
m_k(i)  = \frac{1}{T}\sum \limits_{t=1}^T (R^{\prime}_{i}(t))^k \label{moment}
\end{equation}
For example, the second moment gives the variance as
\begin{eqnarray}
\sigma(i) = \frac{1}{T}\sum \limits_{t=1}^T (R^{\prime}_{i}(t))^2 
\end{eqnarray}
These definitions are used in the analysis subsequently.

Our paper is organized as follows. In section \ref{auto} we explore the individual sectors mentioned above and use the
data to determine some features of the distribution of the returns and find significant deviations from normality.
We then calculate the autocorrelation of log returns for all sectors indices, to test the market efficiency, and find that 
there is significant autocorrelation in most of the sectors of BSE  at lag 1. The more surprising result is that the analysis of our data shows that the
autocorrelations in some sectors persist at higher lags. In section \ref{cross} we analyze the cross-correlations among sectors in BSE. 
Our study spans over FY 2006 - 2013, a time span which consisted a period large fluctuation in indices movement and normal fluctuation period. 
We find that, almost all sectors are highly correlated during period 2006 - 2009 and they are moderately correlated during 2009 - 2013. 
We finally conclude in section \ref{conclude} with a summary of our results and its interpretations. 

%%%%%%%%%%%%%%%%%%%%%%%%%%%%%%%%%%%%%%%%%%%%%%%%%%%%%%%%%%%%%%%%%%%%%%%%%%%%%%%%%%%%%%%%%%%%%%%%%%%%%%%%%%%%%%%%%%%%%%%%%%%%%%%%%%%%%%%%%%%%%%%%%%%%%%%%%%%%%%%
\section{Understanding BSE sectors} \label{auto}

It is commonly believed that the distribution for log return of a stock or for log change in a index
movement is a normal distribution. However, many empirical studies shows deviation from this perception. 
Consequently, any  prediction based on the normal distribution will generally fail. 
In particular, if there is any deviation from normality, the Random Walk Hypothesis will
not be valid. Therefore, it is essential to first understand the distribution of any stock or index movement before 
using any model. Let us first consider the distribution of returns for the various sectors.

The study of Skewness and Kurtosis is very useful to characterize the distribution. 
We know that if a distribution is normal, then sample Skewness and sample excess Kurtosis will be close to zero 
\cite{omnibustest}. Any significant deviation from zero indicates a deviation from normality. 
\begin{figure}
\centering
\subfigure{
\includegraphics[height=3in,width=3in,angle=0]{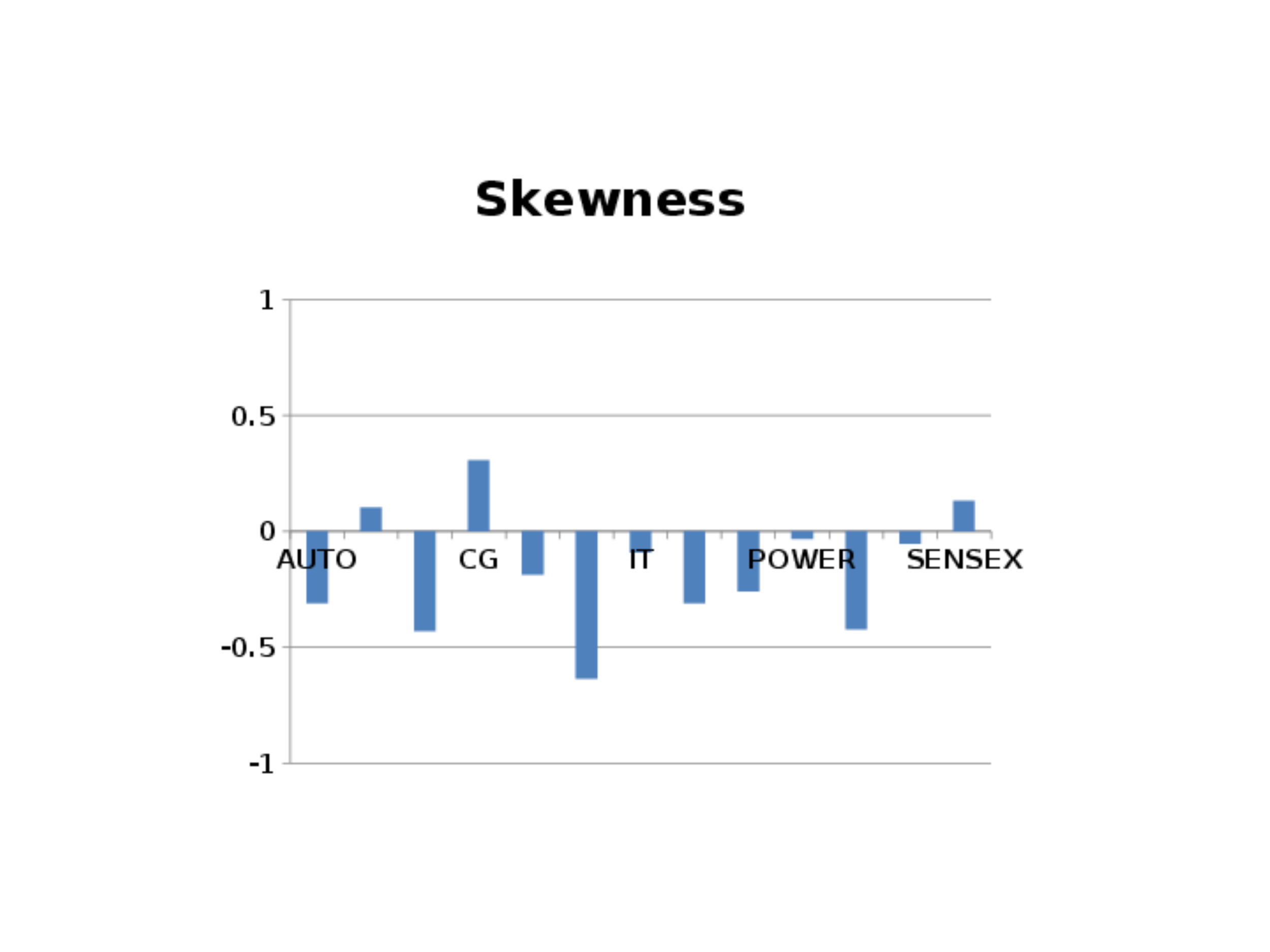}}  
\quad
\subfigure{
\includegraphics[height=3in,width=3in,angle=0]{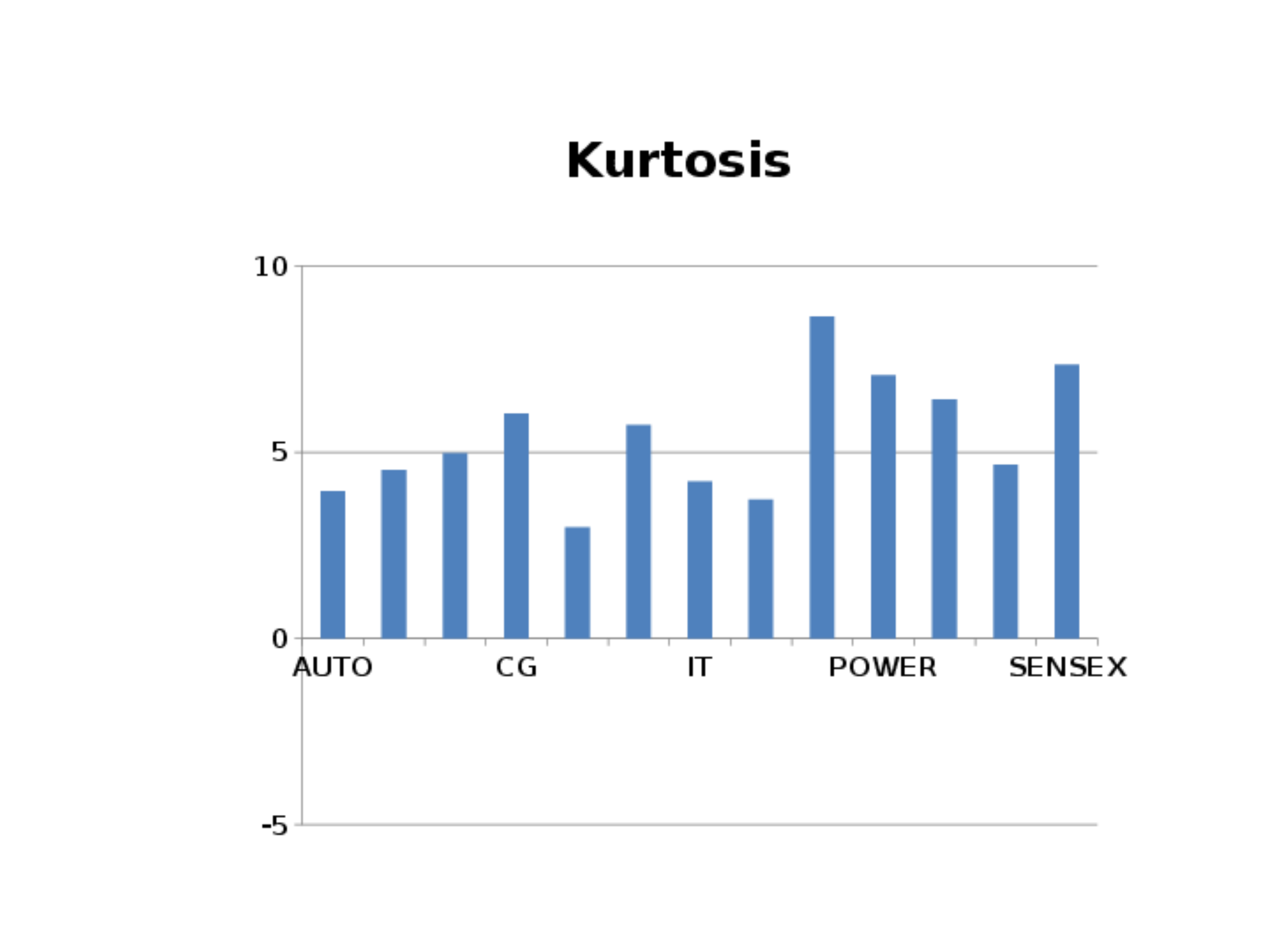}}  
\caption{Sample Skewness and Sample excess Kurtosis of all the Sectors for April 2006-Mar 2014}
\label{figure2}
\end{figure}
The sample skewness and excess kurtosis of any distribution of the $i$th sector can be written in terms of the moments (\ref{moment}) as 
\begin{eqnarray}
\mbox{Sample Skewness:}&&\hspace{2em}  G_{1}(i)  = \frac{T(T-1)}{(T-2)} \frac{m_3(i)}{{m_2(i)}^{3/2}} \label{skewness} \\
\mbox{Sample excess Kurtosis:} &&\hspace{2em} G_{2}(i)  = \frac{(T-1)}{(T-2)(T-3)} 
\left[(T+1)\left(\frac{m_4(i)}{{m_2(i)}^{2}}-3\right)+6 \right] \label{kurtosis}
\end{eqnarray}
The Standard Error in Skewness(SES) and Standard Error in Kurtosis(SEK) are given by\cite{omnibustest}
\begin{eqnarray}
SES = \sqrt{\frac{6T(T-1)}{(T-2)(T+1)(T+3)}} \hspace{2em};\hspace{2em}
SEK = 2 SES \sqrt{\frac{T^2-1}{(T-3)(T+5)}} \label{stnderr}
\end{eqnarray}

The sample skewness and sample excess kurtosis of the above sectors are displayed in Figure \ref{figure2}. 
The Standard Error in Skewness($SES=0.06$) and Standard Error in Kurtosis($SEK=0.11$) are calculated based on the
formulae given by (\ref{stnderr}). 
It is clear from  Figure \ref{figure2} that there is significant deviation from zero for sample Skewness and sample excess Kurtosis in all sectors. 
Hence, based on the study of sample Skewness and sample excess Kurtosis, we can say confidently that each individual
sector's return shows positive  Kurtosis (fat tails) accompanied by Skewness. This clearly shows that the returns of
none of the sectors are normally distributed. 

To further strengthen this claim we perform the D'Agostino-Pearson omnibus test \cite{omnibustest}. 
It is based on two quantities depending on both  Skewness and Kurtosis. The quantities are defined as follows:
\begin{eqnarray}
 Z_{G1}(i) & = &\frac{G1}{SES} \hspace{2em};\hspace{2em} Z_{G2}(i) = \frac{G2}{SEK} \hspace{2em}\hspace{2em} \\
 DP(i) & =& (Z_{G1})^2 +(Z_{G2})^2 
\label{zdef}
\end{eqnarray}

If the distribution of $i_{th}$  sector is normal, the $DP(i)$ should be that of $\chi$ distribution. 
So if $DP(i) > \chi_{critical}^2$,  then the distribution of the $i_{th}$  sector is not normal. For a normal distribution of  
$i_{th}$  sector, $\chi_{critical}^2(2df)$ should be   $13.82$  
with significance level of $0.1\%$. What we find is that the $DP$ values for all the sectors are  much larger than $13.82$. 
Therefore, the statistical results clearly indicate that the data does not satisfy the normality
assumption, i.e. the change in index movement of individual sectors shows  large deviation from normal distribution . 
This finding is also consistent with recent works \cite{devnorm} and shows that returns are driven by assymetric and 
fat-tailed distribution. This also clearly indicates that the market cannot be modeled using the Random Walk Hypothesis \cite{Fama}. 
For stock market modeling or from the perspective of portfolio management the mean-variance model \cite{Mark} should be 
expanded by mean-variance-skewness-kurtosis based portfolio optimization \cite{Lai}. 

To further explore the nature of the auto correlation, we look at the time series of the auto correlation data for the
various sectors. This study is important because if there is autocorrelation in the time series we can  predict immediate 
future based on present information. If there is no autocorrelation in the time series data, 
the data are uncorrelated and it is not possible to make future predictions confidently.

To emphasize, if there is autocorrelation in the time series at lag $1$, 
it is possible to make predictions about immediate future with high degree of certainty. 
Here, we have estimated the sample auto covariance  at lag $k$ for a finite $i_{th}$ time series 
$R_{i}(t)$ of T observations by \cite{BOX} 
\begin{equation}
\gamma_{i k}  = \frac{1}{T}\sum \limits_{t=1}^{T-k} (R_{i}(t)-\bar{R_{i}})(R_{i}(t+k)-\bar{R_{i}}) \label{autocov}
\end{equation}
where $R_i(t)$ is given by our definition (\ref{return}). The autocorrelation at lag $k$ can then be estimated as:
\begin{equation}
\rho_{i k}  = \frac{\gamma_{i k}}{\gamma_{i 0}} \label{autocor}
\end{equation}
The function $\rho_{i k}$ is known as the Auto Correlation Function(ACF).

We have used the Bartlett's approximation \cite{bartlett} to estimate  the variance of the ACF, 
at lags $ k$ greater than some value $ q$ beyond which the  autocorrelation function may be deemed to have 
\textit{died out}. This is defined as \cite{BOX}:
\begin{equation}
var [\rho_{i k}]  \approx  \frac{1}{T} \sum \limits_{\nu=1}^{q}(1+2\rho^2_{i \nu}) \hspace{3em}  k>q
\end{equation}
The standard error  for estimated autocorrelation $\rho_{i k}$ is:
\begin{equation}
SE[\rho_{i k}]=\sqrt{var[\rho_{i k}]}  
\end{equation}
\begin{figure}
\centering
\subfigure[ACF for FMCG]{
\includegraphics[height=3in,width=3in,angle=0]{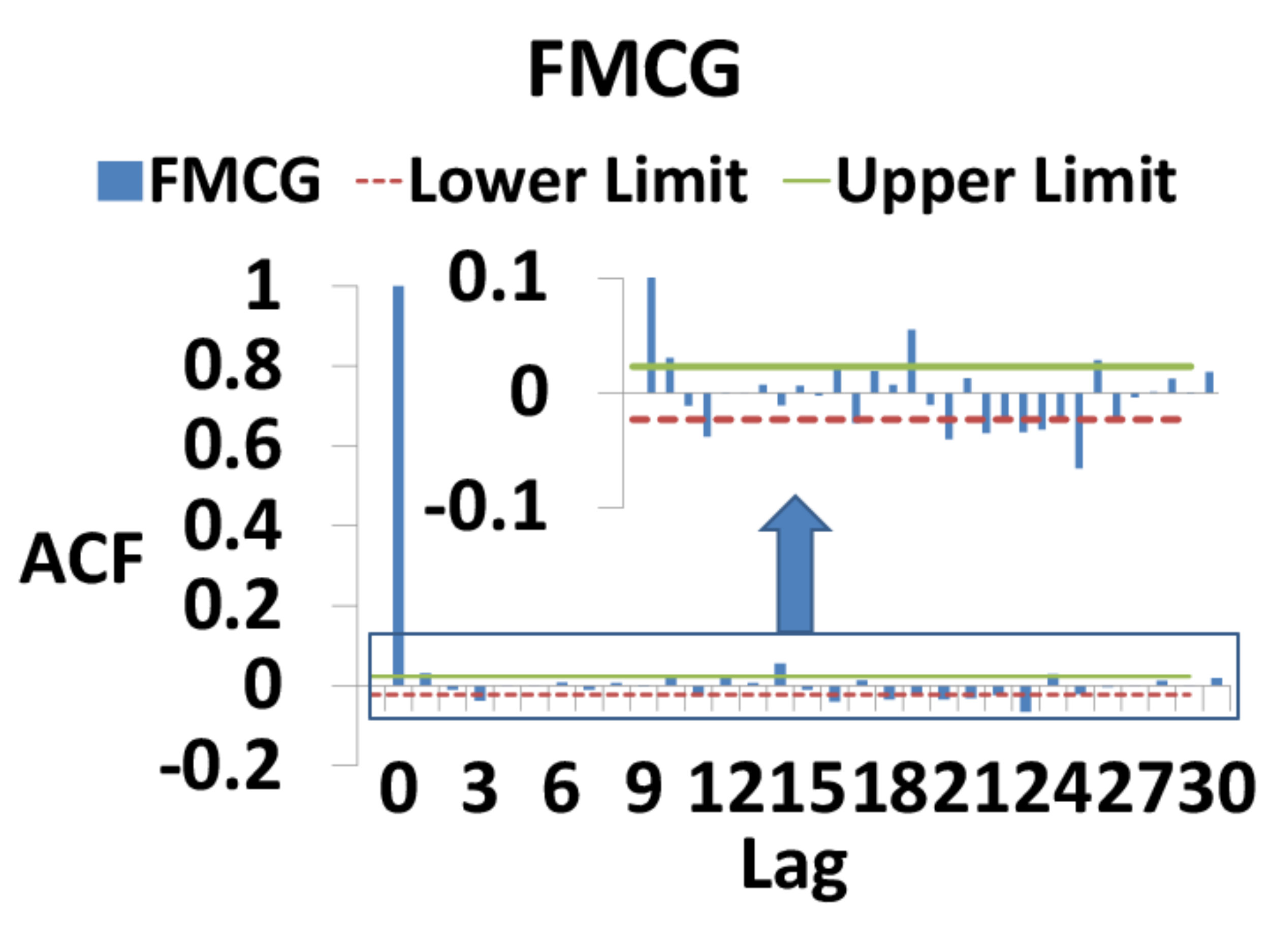}}  
\quad
\subfigure[ACF for Sensex]{
\includegraphics[height=3in,width=3in,angle=0]{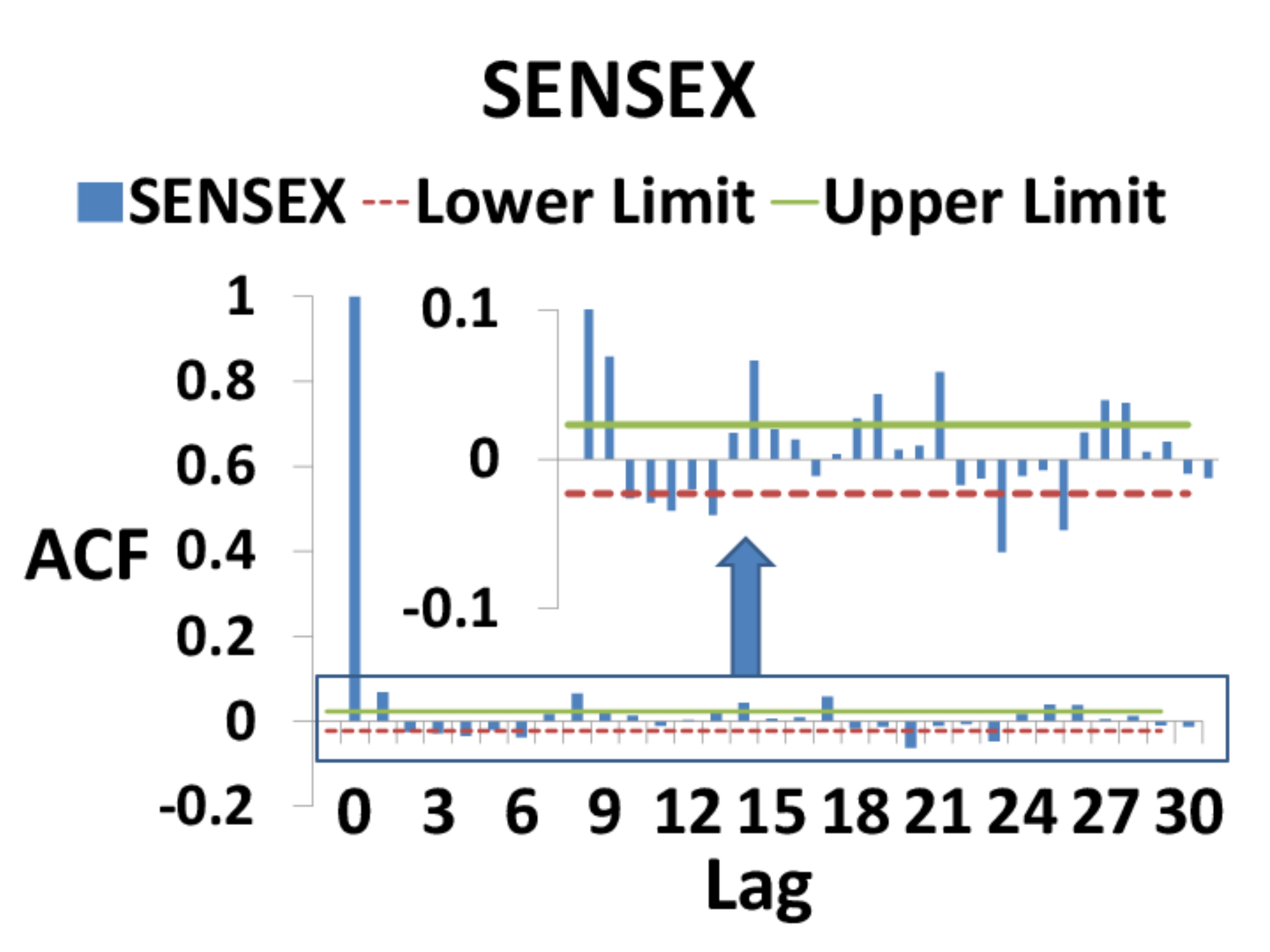}}  
\quad
\subfigure[ACF for Realty]{
\includegraphics[height=3in,width=3in,angle=0]{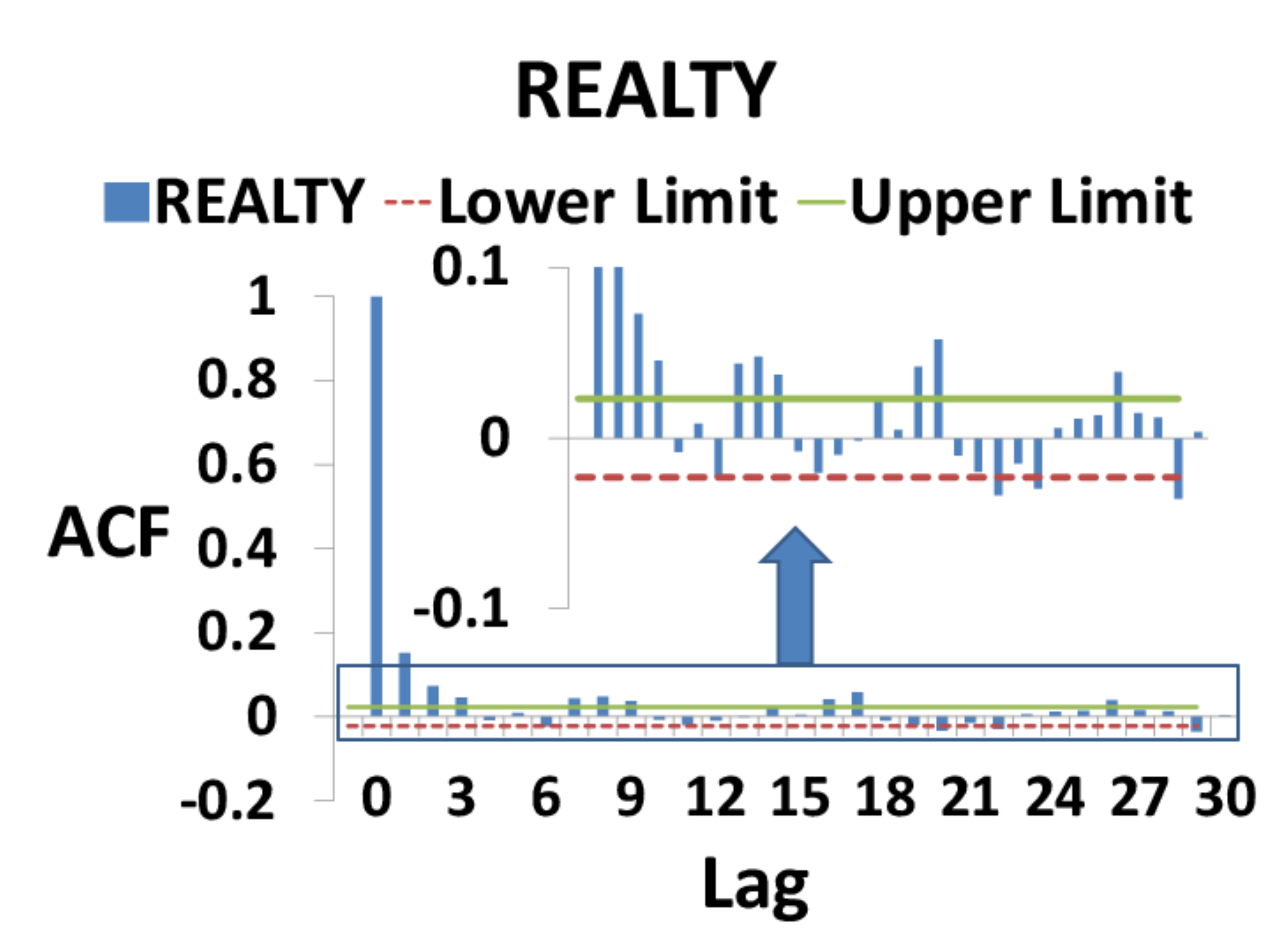}} 
\caption{ACF for FMCG, BSE SENSEX and REALTY sectors for April 2006-Mar 2014}
\label{figure3}
\end{figure}

We calculate the auto correlation of log returns for all sector indices  of BSE.  
It clearly shows that there is significant  autocorrelation in most of the sectors of BSE  at lag 1. 
Therefore, residual effect is confirmed in almost all sectors in BSE. A statistically significant ACF 
value at lag 1 indicates an autoregressive component exists in the time series. In fact, we find some auto correlation 
in most of the  sector persists over time. 

Our results show that there is very little auto correlation in FMCG, weak autocorrelation in 
IT, Teck and Oil and Gas, and significant lag $1$ auto correlation in Auto, Bankex, CD, CG, HC,  Metal,  Power, Realty, and also in Sensex. 
In Figure \ref{figure3}, we have  plotted the ACF for three BSE sectors for illustration. The figure also shows
how the ACF persists in time. This feature, obtained by the analysis of our data, is extremely striking and has not been
reported in literature before. Further analysis is required, in future works, to fully understand this feature.    

The study of the ACF is an empirical test of the efficiency of the BSE market for the period under consideration. 
The persistence of auto correlation we see above clearly indicates that the BSE is not an efficient market. For example
Figure \ref{figure3}  shows the significant consistent autocorrelation in REALTY (lags $1,2,3,7,8,9,16,17,26$), 
FMCG  (lags $14,23$) and Sensex (lags $1,8,14,17,20,23$). Without high frequency data it is not possible to comment why
these exact days lag are significant. However this broad analysis shows that during the period under consideration the
BSE was not even weak efficient. 
 
According to the Efficient Market Hypothesis(EMH), the stock prices fully reflect any changes in the information
available to investors. For example, a market following random walk is consistent with the EMH. It has been shown \cite{efficient} 
that mature stock markets are generally \textit{weak} efficient. A departure from weak efficiency (i.e. deviation from random walk)
may point towards possible market manipulation.  

The autocorrelation exhibited by the BSE sectors agrees with the 
findings in \cite{LO}. Those authors also  show that autocorrelation in returns might generate a momentum. 
Therefore, a BSE sector that outperformed other sectors in the past might continue to do so for some time interval. 
These features in the auto correlation may be crucial for portfolio management in Indian equity markets
 
\begin{figure}
\centering
\includegraphics[height=3in,width=5in,angle=0]{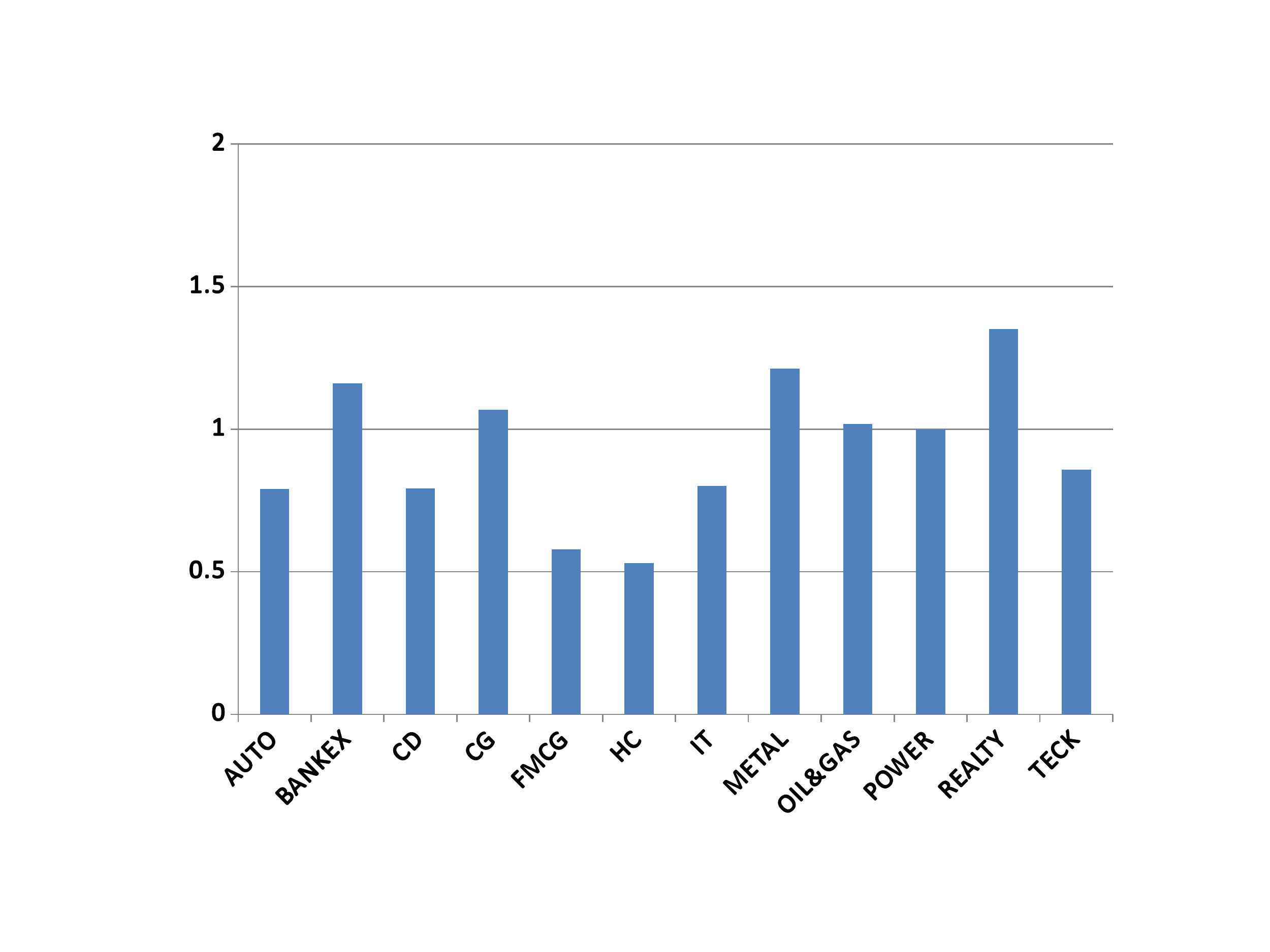}  
\caption{$\beta$ of all the Sectors for April 2006-Mar 2014}
\label{figure4}
\end{figure}

Financial market volatility is central to the theory and practice of asset pricing, asset allocation, and risk
management. Popular assumption is that volatilities and correlations are constant, but we have seen that they
have significant variation over time. Therefore, the study of $\beta$  can be useful for investor \cite{Beta}. The $\beta$ factor is defined as 
\begin{equation}
\beta_{i}=\frac{\mbox{covariance}(i, Sensex)}{\mbox{variance}(Sensex)} 
\end{equation}
A $\beta$ of 1 indicates that the security's price will move with the market. A $\beta$ of less than 1 means that the security will be less volatile 
than the market. A $\beta$ of greater than 1 indicates that the security's price will be more volatile than the market.
As an example, from Figure \ref{figure4} we can see that the $\beta$ of the Bankex sector is 1.45 ( 45$\%$ more volatile than the market) 
while that of the HC sector is 0.49 ( less volatile than the market). A systematic study of this parameter will be
undertaken in a future work.

%

%%%%%%%%%%%%%%%%%%%%%%%%%%%%%%%%%%%%%%%%%%%%%%%%%%%%%%%%%%%%%%%%%%%%%%%%%%%%%%%%%%%%%%%%%%%%%%%%%%%%%%%%%%%%%%%%%%%%%%%%%%%%%%%%%%%%%%%
\section{Cross-correlation among BSE sectors} \label{cross}
In the last section we have shown that there is significant autocorrelation in most  sectors. Let
us now try to see whether the movement of the indices in various sectors are also correlated i.e. whether there exists any
cross correlation between the sectors. Some study of cross correlations of other markets have been carried out in
different contexts \cite{collec} but to the best of our knowledge, there exists no studies of the correlations between
sectors at least in the context of the Indian financial markets.

To understand the interactions among the sectors, it is useful to study the spectral properties of the
correlation matrix of sectorial indices movements. The deviation of eigenvalues of the correlation correlation matrix
from those of a random matrix provide signals about the underlying interactions between various sectors.  
The largest eigenvalue is identified as representing the influence of the entire market,
common for all sectors. The remaining large eigenvalues are associated with the different  sectors, as indicated by the composition of
their corresponding eigenvectors \cite{Gopikrishnan}. This is what we do in this subsequently in this section. 

\begin{figure}
\centering
\includegraphics[height=5in,width=5in,angle=0]{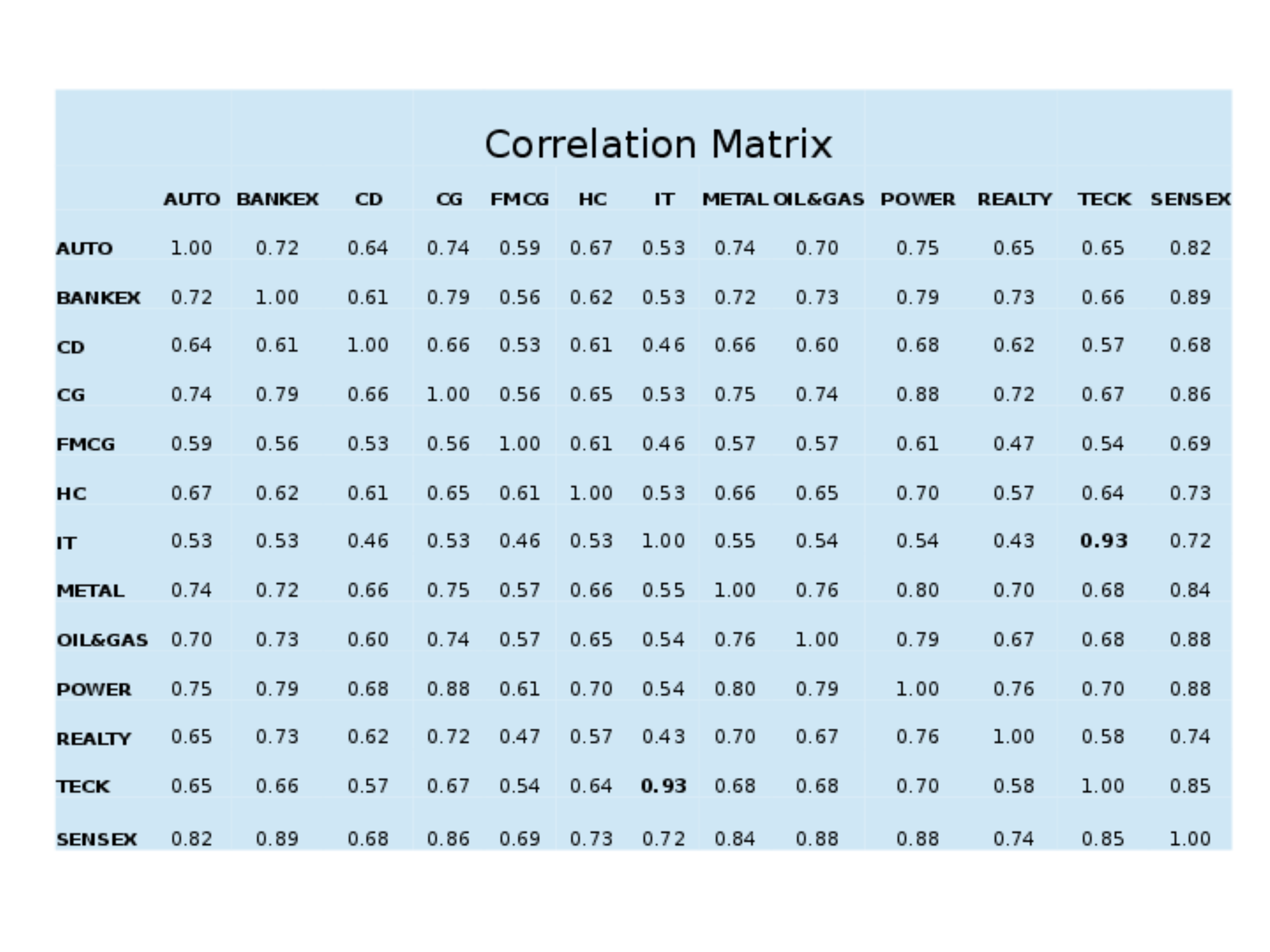} 
\caption{Cross-Correlation matrix for April 2006-Mar 2014}
\label{correl.matrix}
\end{figure}

\begin{figure}
\centering
\includegraphics[height=5in,width=5in,angle=0]{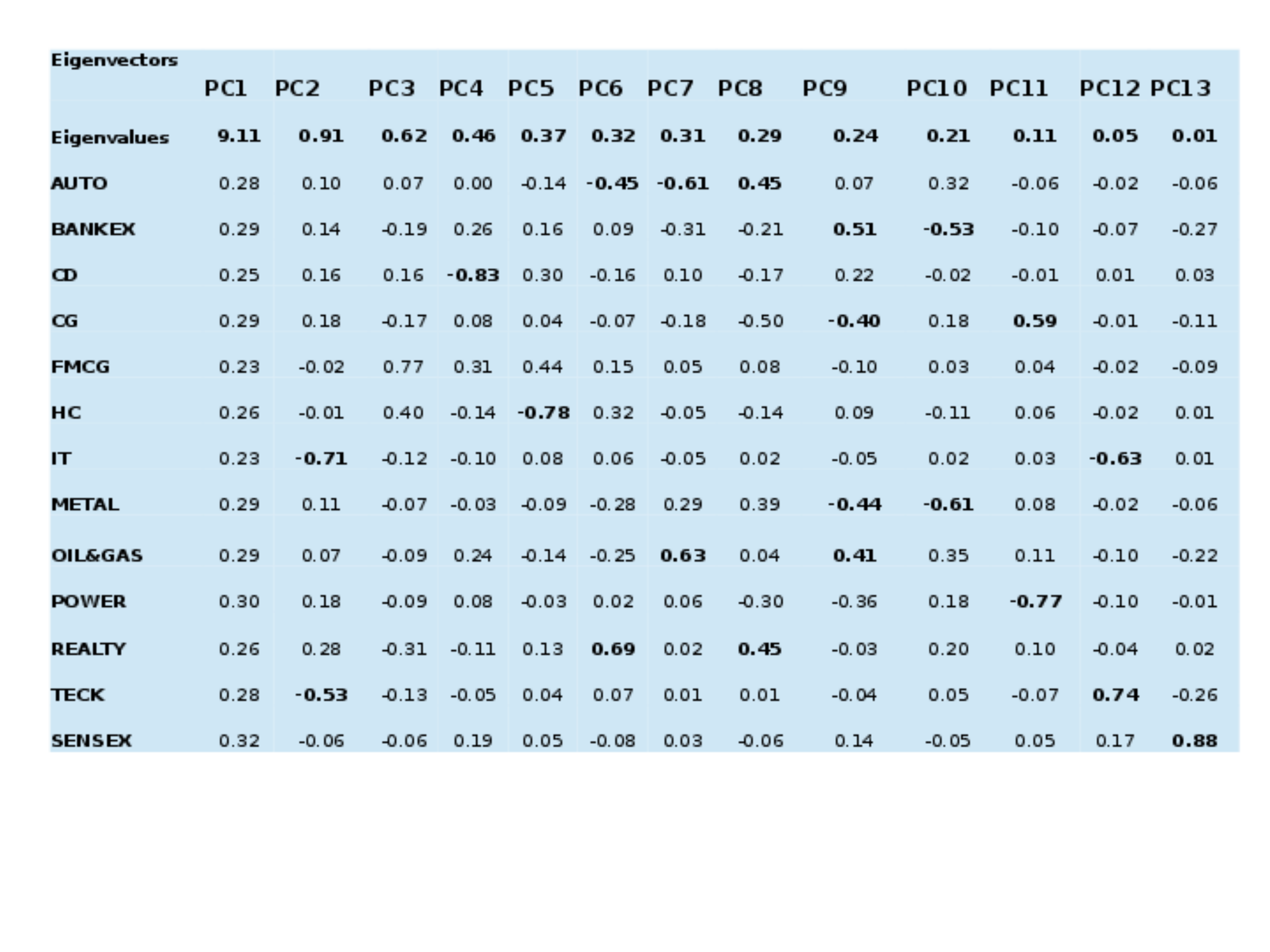}  
\caption{Eigenvalues and Eigenvectors of Correlation matrix for April 2006-Mar 2014}
\label{figure6}
\end{figure}

If the time series of returns of $N$ sectors of length $T$ are mutually uncorrelated, then
the resulting correlation matrix is random and is known as Wishart matrix \cite{MarchPast}. 
It is known that the empirical distribution of the eigenvalues of the Wishart matrix almost always converges to a 
probability distribution as $T\rightarrow\infty$ and $\frac{N}{T}\rightarrow a$ where $a$ is a constant such that $0 \leq a < 1$. 
In that limit the distribution is continuous and supported on $\lambda \in [ (1-\sqrt{a})^2,(1+\sqrt{a})^2]$ where 
$0 < a < 1$ \cite{MarchPast}. This bound is known as the Random Matrix Theory (RMT) bound. 
Therefore, the eigenvalue $\lambda$ of the Wishart matrix should lie  between $0.84$ and  $1.17$. 
We estimate the sample cross correlation matrix for our data set i.e for $N=13$ sectors for $T=1990$ days and
$\frac{N}{T}\rightarrow a=0.006533$ (see Figure \ref{correl.matrix}).
The reduced number of Principal Components (PC) of the cross correlation matrix that can explain most of the total variance
is given in terms of the eigenvectors ${\mathbf {u}}$ of the cross correlation matrix as
\begin{equation}
{\mathbf C}= \sum_{i=1}^{N}\lambda_i{\mathbf u}_i{\mathbf u}_i^T,
\end{equation}
We find the eigenvalues $\lambda_i$ of the cross correlation matrix. The eigenvectors corresponding to these
eigenvalues are the PCs of the cross correlation matrix. These eigenvectors can be expanded in a basis given by the 13 sectors we
are considering. All the eigenvalues and the expansion of the PCs in our chosen basis is given in Figure
\ref{figure6}. 

From Figure \ref{figure6} we can see significant deviation of the largest eigenvalue of the PC1 from the
largest eigenvalue of RMT. The largest eigenvalue of the cross correlation matrix is 9.11. 
Also, from the first column (corresponding to PC1) of Figure \ref{figure6}.
the eigenvector of largest eigenvalue shows a relatively uniform composition, i.e. all sectors 
contribute to it and all elements having the same sign. 

\begin{figure}
\centering
\includegraphics[height=3in,width=5in,angle=0]{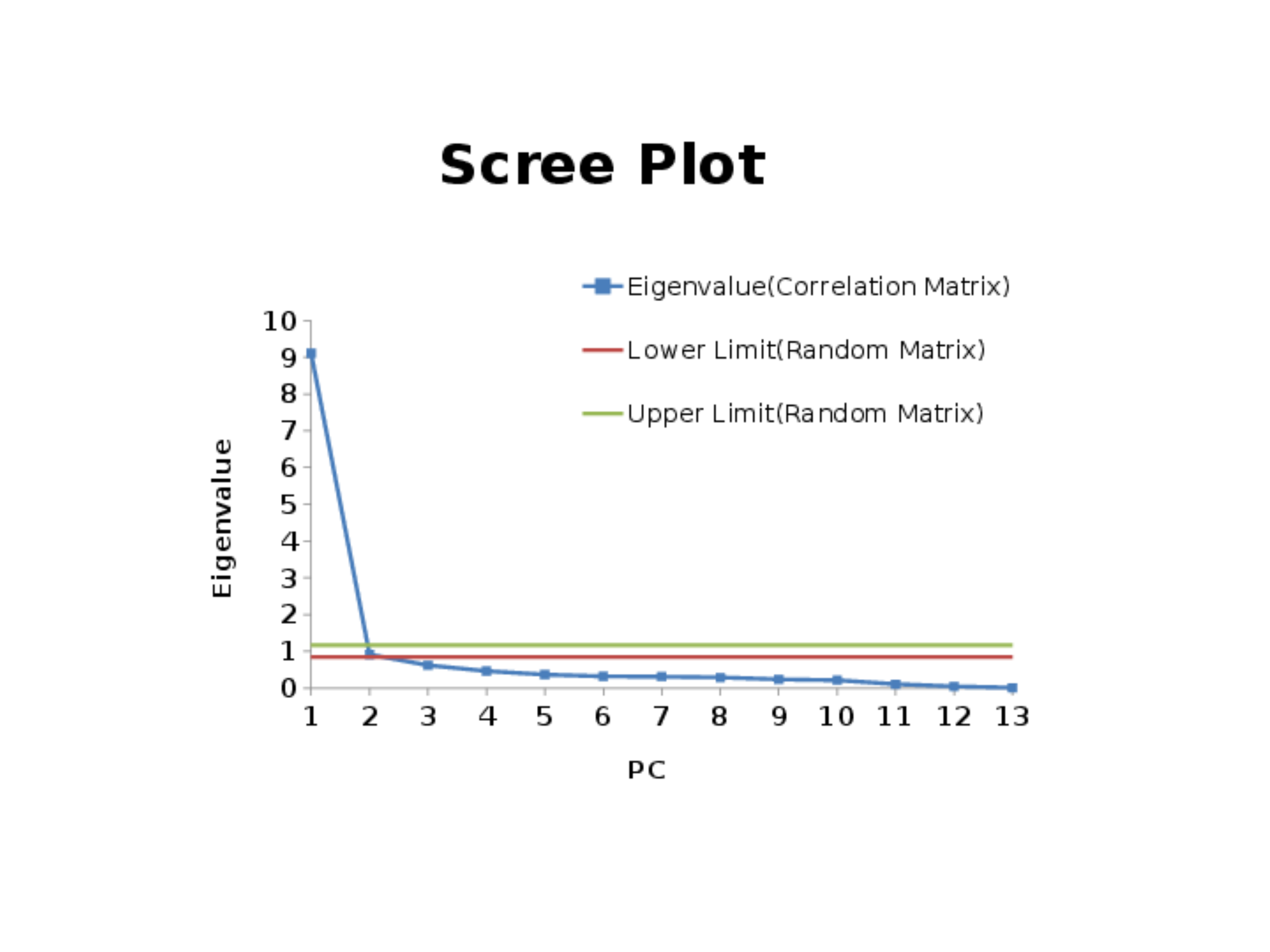}  
\caption{Eigenvalue and Eigenvector matrix of Correlation matrix for April 2006-Mar 2014 and Scree Plot for April 2006-Mar 2014}
\label{figure7}
\end{figure}

A very useful visualization of what we discussed above is the scree plot \cite{scree} as can be seen in Figure
\ref{figure7}. The fact that the PC1 is so large and that it affects all the sectors with the same
ratio, we can say that the largest eigenvalue is associated with the collective response of the entire market to external
informations \cite{Mantegna99, Laloux99}, i.e. the largest eigenvalue is due to the existence of a
market-induced correlation across all sectors. Since PC1 dominates to such a large extent it is difficult to observe the
correlations between sectors.

From the investment point of view, it is interesting to note that the Tech and the IT  sectors are
highly correlated all the time. Hence, it would be better to club both these the sectors together for modelling and for
portfolio diversification purposes.

\begin{figure}
\centering
\subfigure[April 2008 - March 2009]{
\includegraphics[height=3in,width=3in,angle=0]{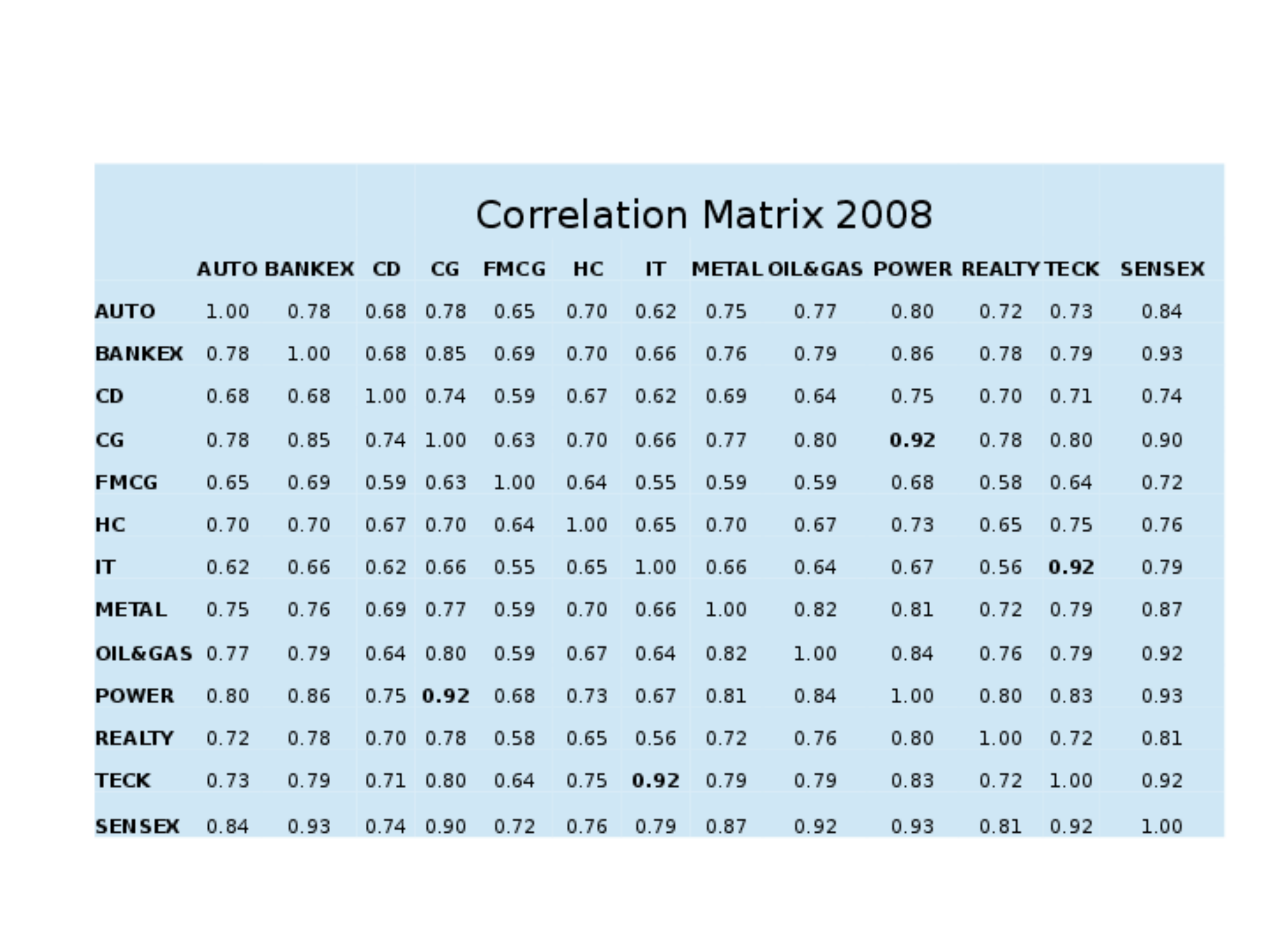}}    
\quad
\subfigure[April 2012 - Mar 2013]{
\includegraphics[height=3in,width=3in,angle=0]{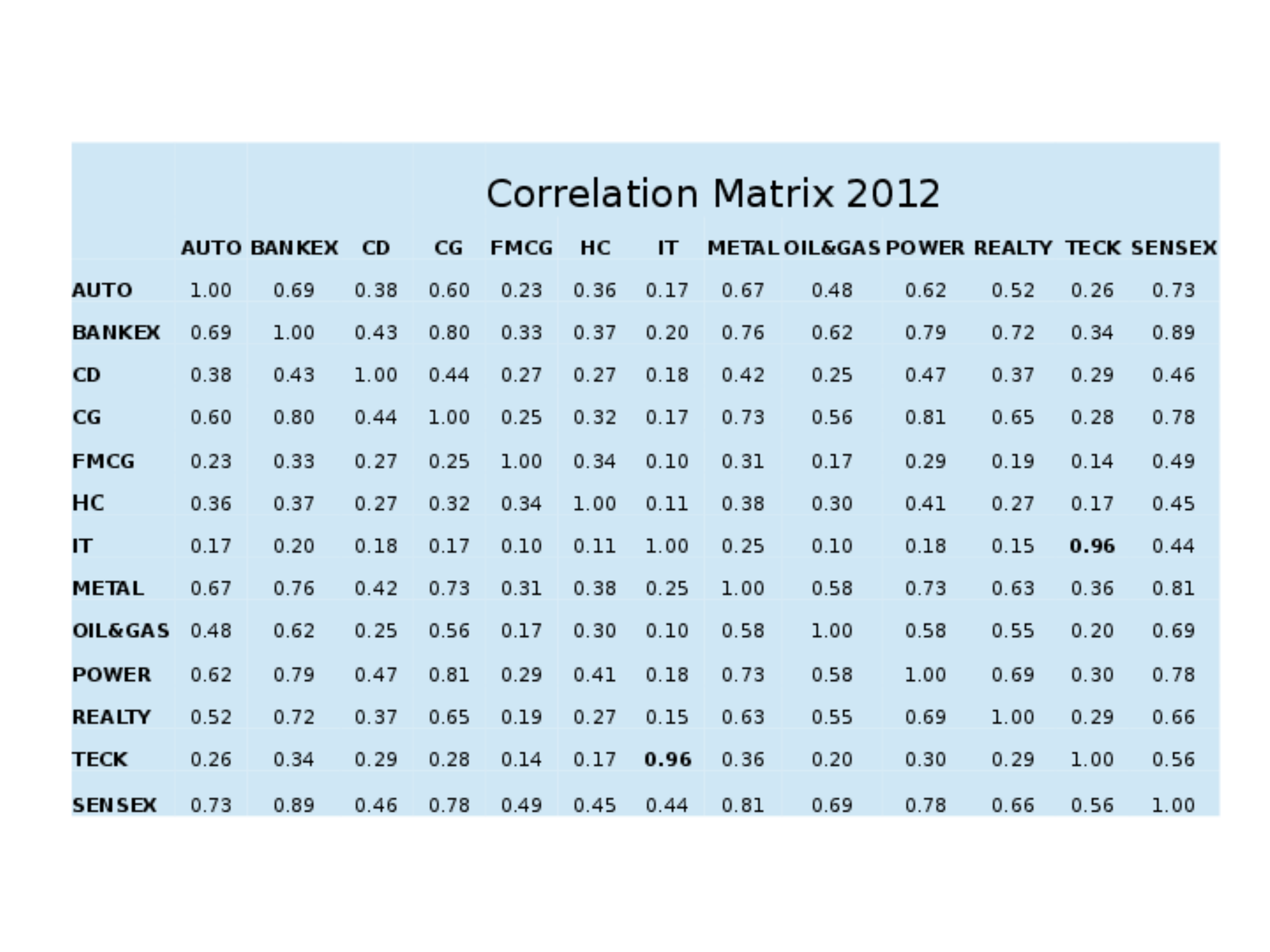}}    
\caption{Comparison of Correlation Matrices}
\label{comparecormat}
\end{figure}

The scree plot also gives some very useful information about periods of large fluctuations. During the time of large
fluctuations we find that there is a is large correlation among most of the sectors. As a comparison consider Figure
\ref{comparecormat} where we compare the cross correlation matrices of a period of large fluctuation (April 2008 - March
2009) with a period of relatively small fluctuation (April 2012 - Mar 2013). As can be seen from the figure, although
there exists significant cross correlations at both the times, the magnitude is lesser in the later period. This
indicates that periods of large fluctuations can be studied using models where the correlation strength becomes large. 
Since periods of large fluctuations may correspond to crashes in the stock market, a systematic
study of the cross correlation matrices of these periods will provide valuable insights into understanding and modeling
crashes.   

\begin{figure}
\centering
\subfigure[April 2008 - March 2009]{
\includegraphics[height=3in,width=3in,angle=0]{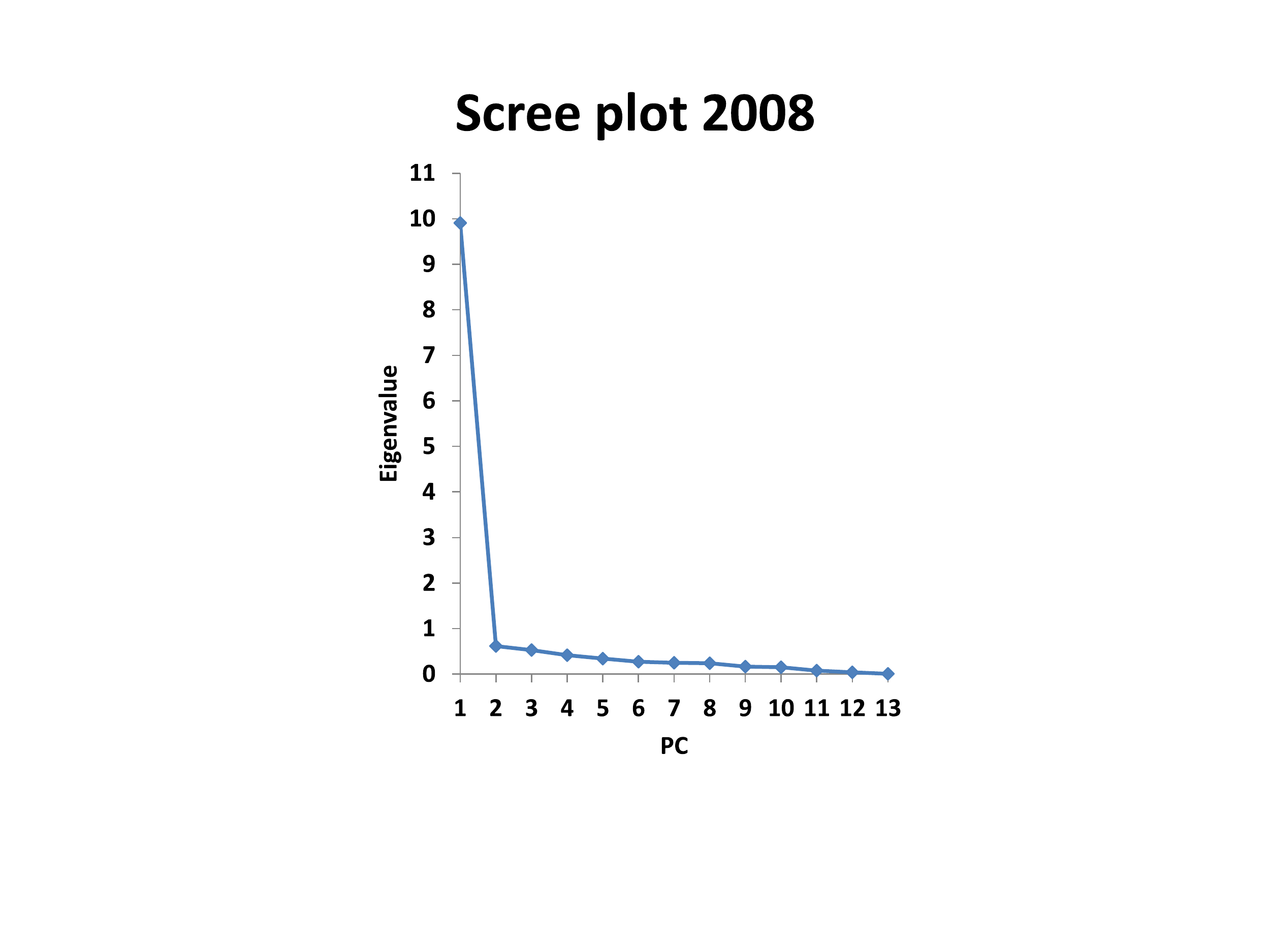}}  
\quad
\subfigure[April 2012 - Mar 2013]{
\includegraphics[height=3in,width=3in,angle=0]{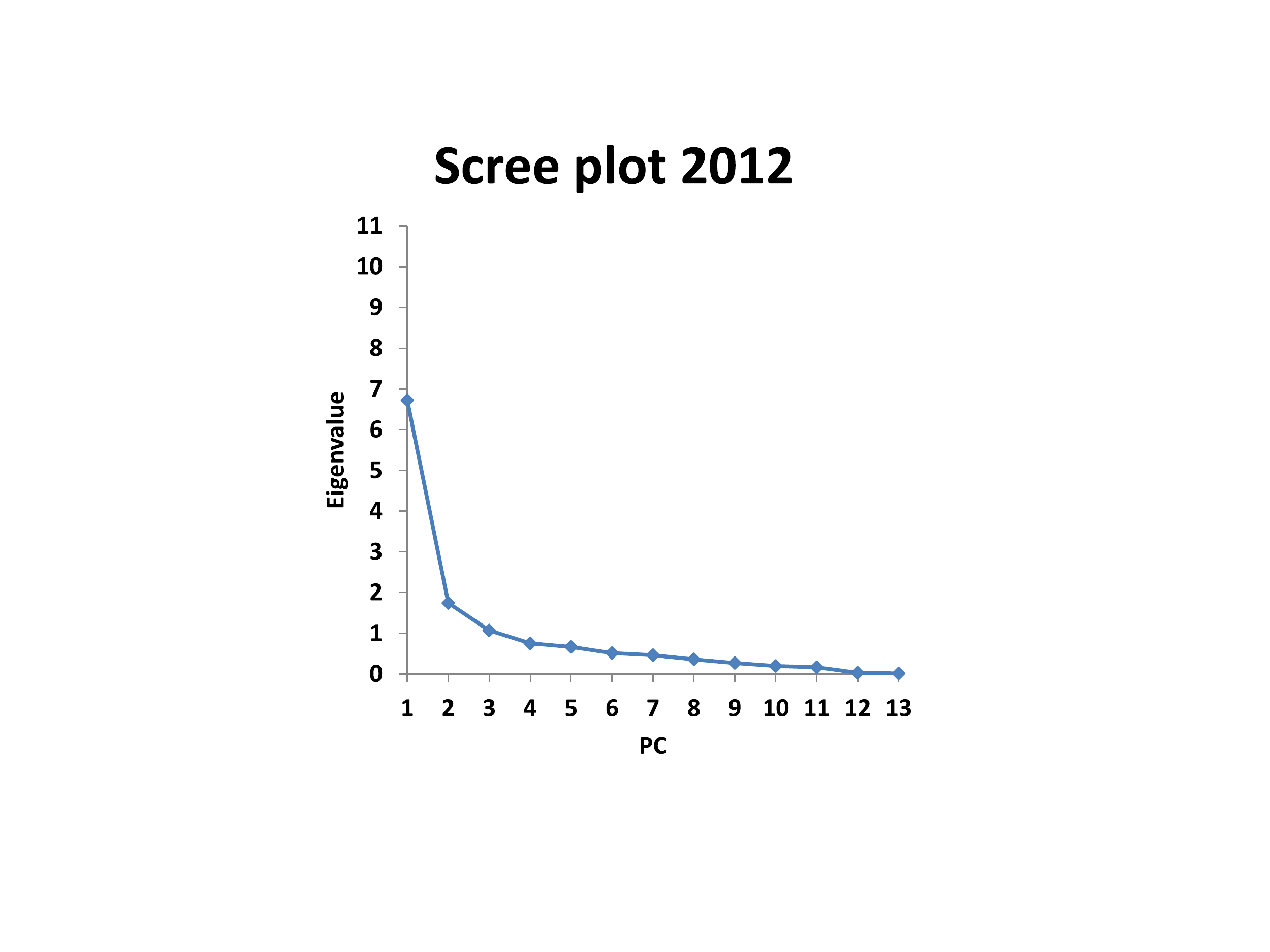}}  
\caption{Comparison of Scree Plots}
\label{comparescree}
\end{figure}

\begin{figure}
\centering
\subfigure{
\includegraphics[height=3in,width=3in,angle=0]{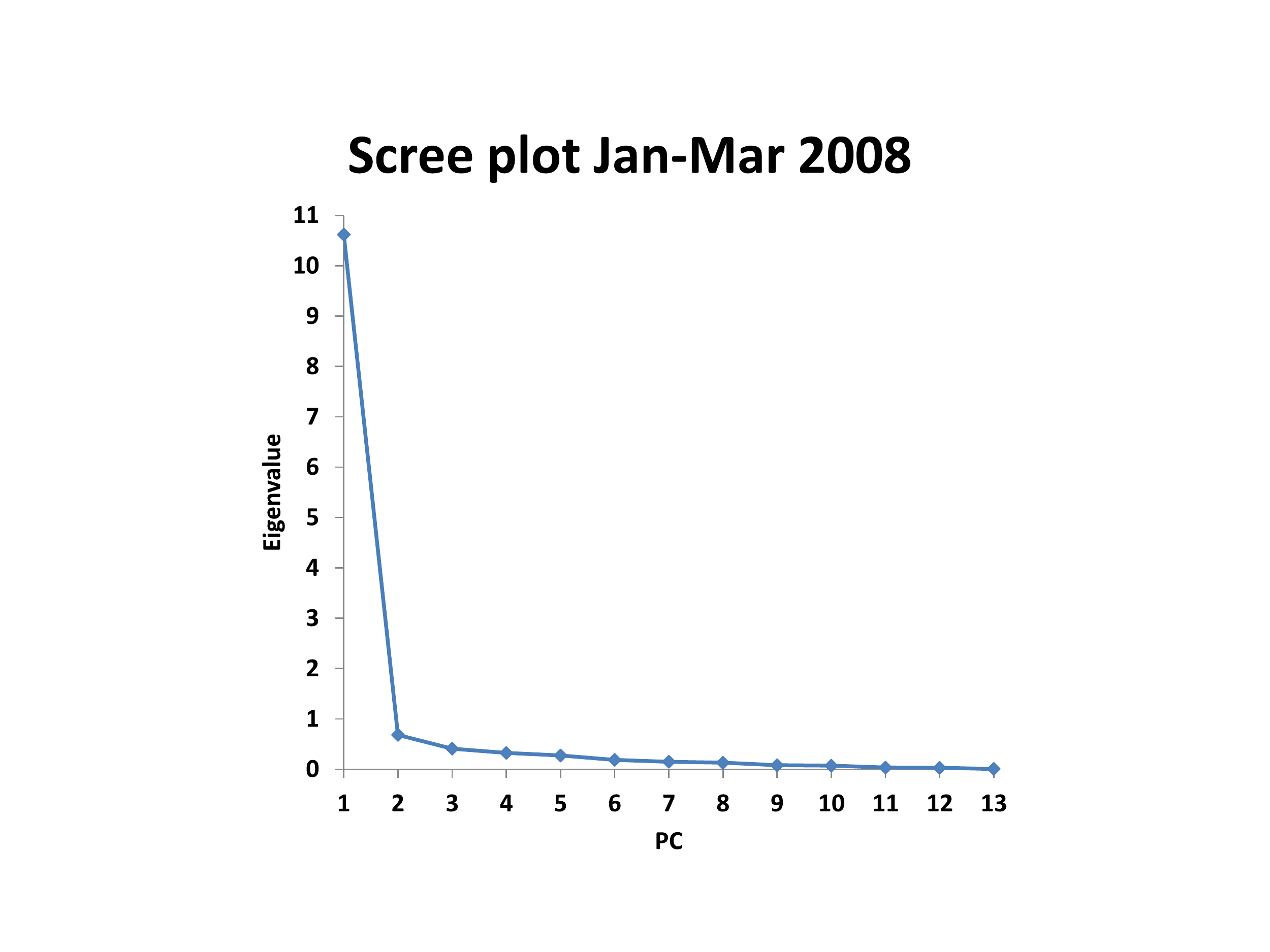}}  
\quad
\subfigure{
\includegraphics[height=3in,width=3in,angle=0]{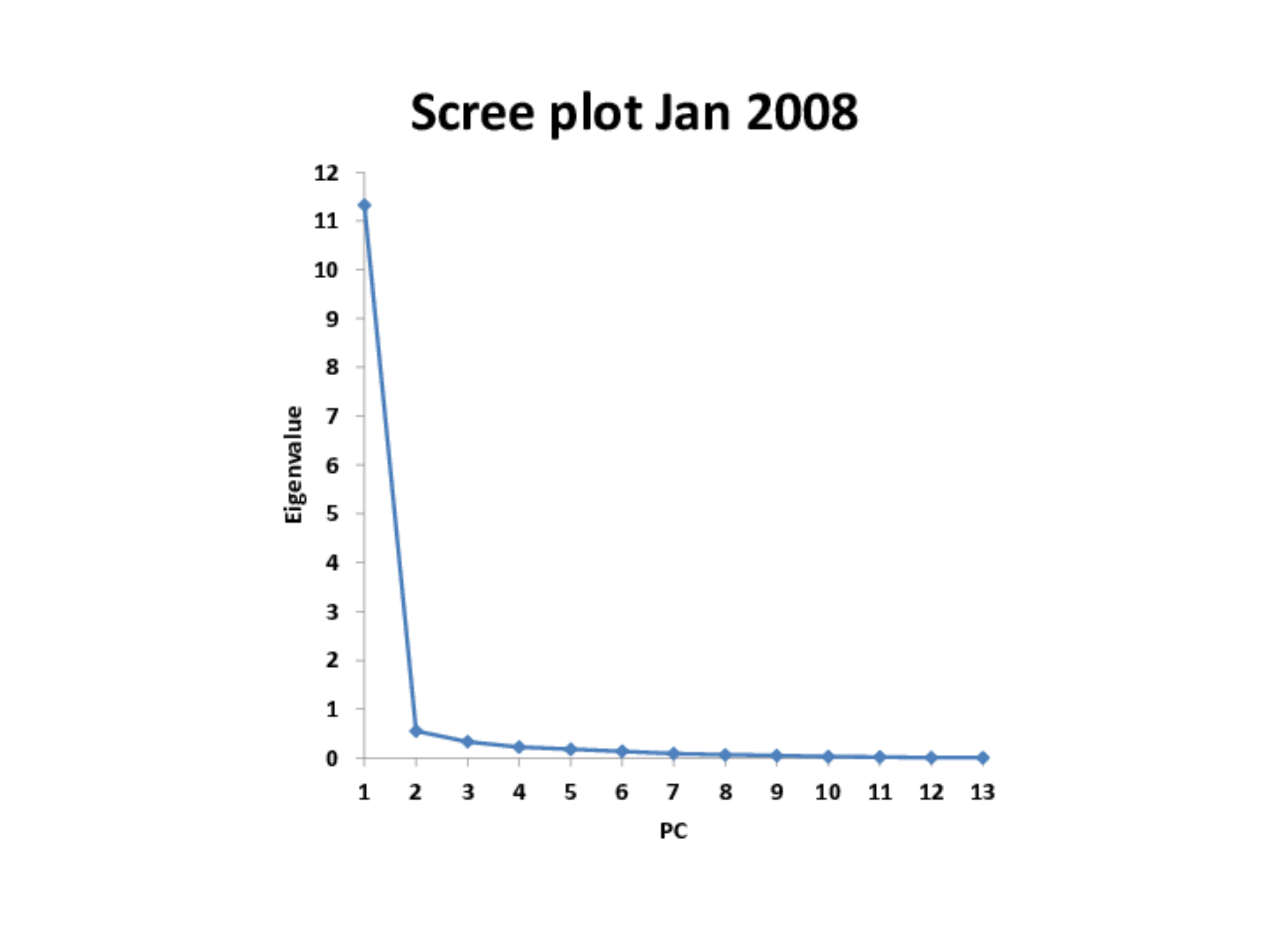}}  
\caption{Scree Plot for April Jan-Mar 2008 and for Jan 2008}
\label{screezoom}
\end{figure}

A more efficient way of analyzing this would be by doing the Principle Component analysis we had performed previously
in this section. Again, scree plots provide a more efficient and rigorous demonstration of the increase in correlations during
periods of crisis. As can be see in Figure \ref{comparescree}, the PC1 when the entire market is
experiencing large fluctuations is 9.91, while it comes down to 6.72 during period of relative calm. We can actually
\textit{zoom in} to the actual time of the crash (Jan 2008) in Figure \ref{screezoom} using the quarterly and monthly
data and see that the PC1 is actually higher (11.32) during that time. This can provide a efficient and novel way of analyzing
crashes of the stock market.
%
%%%%%%%%%%%%%%%%%%%%%%%%%%%%%%%%%%%%%%%%%%%%%%%%%%%%%%%%%%%%%%%%%%%%%%%%%%%%%%%%%%%%%%%%%%%%%%%%%%%%%%%%%%%%%%%%%%%%%%%%%%%%%%%%%%%%%%%%%%%%%%%%%%%

\section{Conclusions} \label{conclude}

In this paper, we have carried out a model independent analysis of the BSE for a period of 1990 days. This time frame
contains periods of both small and large fluctuations and thus provides a good sample to understand and study the
generic behavior of the stock market. Also the number of days chosen was large to avoid small sample size errors.
Instead of studying the movement of individual stock returns as is usually done, we study the movement and behavior of
groups of stocks, the grouping being done in terms of sectors. We look at 12 sectors of stocks and use the whole Sensex
as the benchmark. The auto correlations in the return data captures how the stocks within the individual sectors interact 
among themselves while the cross correlations look at how the sectors affect each other. 

We found the presence of significant auto correlations in all the sectors clearly demonstrating that the movement of the
stock prices cannot be modeled via random walk. While this is usually a accepted feature of stock market model, our
analysis of the departure from normality is rigorous. It is not just based on the non zero skewness and kurtosis but also
on  D'Agostino-Pearson omnibus test. This comprehensively shows the existence of auto correlations. From an investors
point of view, this means that the only mean-variance-skewness-kurtosis based methods of portfolio optimization will be
useful.

A more interesting feature which we find in the study of auto correlations is that they persist over time. 
The ACF is significant for all sectors at lag one and there are certain sectors where this auto correlation 
persists at higher lags. This indicates that the BSE has significant departure from efficient market and EMH cannot be
used to model the stock price movement in BSE. This is a very interesting property of the stock market which has to be accounted for in the future models.
For financial markets to be meaningful and useful to the economy, they must be at least weakly efficient. Some of the
reasons why BSE is not efficient may be (i) weak disclosure procedure (ii) poor quality and quantity of company' disclosure 
(iii) almost no public awareness about securities (iv) no transparent regulation, supervision and administrative rule. 
This feature of the BSE should be of great interest not just to the investors but also policy makers and market
regulators. Further analysis of this will be done in a future work.

 We also study the relative volatility of the sectors compared to the whole market, measured in terms of
$\beta$. This parameter, as we point out, should have a significant role in making investment decisions. How to use this
parameter in building physics models of financial markets is a direction of future work.

The cross correlation was studied by doing the Principle Component analysis of the correlation matrix. Our findings
show that there exist a very large cross correlation but that correlation is due to some external force which drives the 
market as a whole. The effect of sectors on each other is smaller but not insignificant and will be the focus of a
future work. A very important feature following from our analysis is that the value of PC1 increases during periods of
large fluctuation of the market. This can have far reaching application in studying and predicting crashes of financial
markets.

{\bf{Acknowledgement}}

We would like to thank Dr. Sitabhra Sinha for careful reading of the manuscript and helpful suggestions.

%%%%%%%%%%%%%%%%%%%%%%%%%%%%%%%%%%%%%%%%%%%%%%%%%%%%%%%%%%%%%%%%%%%%%%%%%%%%%%%%%%%%%%%%%%%%%%%%%%%%%%%%%%%%%%%%%%%%%%%%%%%%%%%%%%%%%%%%%%%%%%%%%%%%%

\end{document}